%% file: main.tex
\documentclass[10pt,conference]{IEEEtran}
\IEEEoverridecommandlockouts

\usepackage{cite}
\usepackage{amsmath,amssymb,amsfonts}
\usepackage{algorithmic}
\usepackage{graphicx}
\usepackage{textcomp}
\usepackage{xcolor}
\usepackage{booktabs}
\usepackage{url}
\usepackage{comment}
\usepackage{siunitx}

\def\BibTeX{{\rm B\kern-.05em{\sc i\kern-.025em b}\kern-.08em
    T\kern-.1667em\lower.7ex\hbox{E}\kern-.125emX}}
\begin{document}

\title{A Vision-based Framework for \\Intelligent gNodeB Mobility Control
}

\author{
    \IEEEauthorblockN{
        Pedro Duarte, André Coelho, Francisco Ribeiro,
        Filipe B. Teixeira, Luís Pessoa, Manuel Ricardo
    }
    \IEEEauthorblockA{
        INESC TEC, Faculdade de Engenharia, Universidade do Porto, Portugal\\
        \{pedro.r.duarte, andre.f.coelho, francisco.m.ribeiro, filipe.b.teixeira, luis.m.pessoa, manuel.ricardo\}@inesctec.pt
    }
\thanks{This work was supported by the CONVERGE project which has received funding under the European Union’s Horizon Europe research and innovation programme under Grant Agreement No 101094831.}  
}

\maketitle

\input{0_Abstract}

\begin{IEEEkeywords}
O-RAN, reinforcement learning, computer vision, vision-aided networking, multimodal perception.
\end{IEEEkeywords}

\input{1_Introduction}
\input{2_VisionRAN}
\input{3_VisionApp}
\input{4_VisionTwin}
\input{5_Evaluation}
\input{6_Discussion}
\input{7_Related_Work}
\input{8_Conclusions}

\input{9_Bibliography}

\end{document}

%% file: 0_Abstract.tex
\begin{abstract}
This paper proposes a vision-based framework for the intelligent control of mobile Open Radio Access Network (O-RAN) base stations (gNBs) operating in dynamic wireless environments. The framework comprises three innovative components. 
The first is the introduction of novel Service Models (SMs) within a vision-enabled O-RAN architecture, termed \textit{VisionRAN}. These SMs extend state-of-the-art O-RAN-based architectures by enabling the transmission of vision-based sensing data and gNB positioning control messages.
The second is an O-RAN xApp, \textit{VisionApp}, which fuses vision and radio data, and uses this information to control the position of a mobile gNB, using a Deep Q-Network (DQN). 
The third is a digital twin environment, \textit{VisionTwin}, which incorporates vision data and can emulate realistic wireless scenarios; this digital twin was used to train the DQN running in \textit{VisionApp} and validate the overall system. 
Experimental results, obtained using real vision data and an emulated radio, demonstrate that the proposed approach reduces the duration of Line-of-Sight (LoS) blockages by up to 75\% compared to a static gNB. These findings confirm the viability of integrating multimodal perception and learning-based control within RANs.
\end{abstract}

%% file: 1_Introduction.tex
\section{Introduction}\label{chap:1}

Traditional mobile networks have been built on monolithic and proprietary architectures that restrict flexibility, hinder interoperability, and limit the pace of innovation. To address these challenges and meet the growing demand for wireless connectivity, cellular networks are increasingly adopting open and modular architectures. A representative example is the Open Radio Access Network (O-RAN) architecture~\cite{oran}, which defines standardized interfaces, such as the E2 interface, and introduces near-real-time Radio Access Network Intelligent Controllers (near-RT RICs) that enable data-driven control through modular software applications known as xApps.

With these advances, future Radio Access Networks (RANs) are expected to operate in increasingly dynamic environments, where factors such as user mobility and line-of-sight (LoS) obstructions present significant challenges to wireless link reliability and overall network performance. These conditions underscore the need for mobile base stations (gNBs) capable of autonomous repositioning in response to environmental changes. In parallel, the integration of perception-aided systems offers the potential to enhance environmental awareness and support more informed control decisions.

This paper was conducted within the Horizon Europe CONVERGE project~\cite{converge_web}, which aligns with O-RAN principles and aims to unify radio and vision systems to enable mobile networks that ``view to communicate and communicate to view''. While O-RAN promotes openness and programmability, its native interfaces primarily convey conventional radio and network metrics, such as Signal-to-Noise Ratio (SNR) and traffic statistics. Although these metrics are essential for network management, they provide limited knowledge of the physical environment and lack the contextual awareness necessary for perception-driven control.

Future sixth-generation (6G) networks are expected to incorporate context-aware, predictive control mechanisms to meet increasingly stringent performance and reliability requirements. Achieving this vision may benefit from the integration of external sensing modalities, such as vision and positioning. However, the current O-RAN specification lacks native support for such data, revealing a gap between existing architectures and the requirements of emerging intelligent network controllers.

The main contribution of this paper is a \textbf{vision-based framework for the intelligent control of mobile O-RAN gNBs} operating in dynamic wireless environments. The framework comprises \textbf{three innovative components}. First, \textbf{new Service Model (SM) messages}, generated by dedicated E2 agents, extend the E2 interface to support the transmission of positioning and visual data. These messages are implemented within \textit{VisionRAN}, a vision-enabled extension of the CONVERGE architecture \cite{teixeira2025converge}. Second, a \textbf{vision-aided O-RAN xApp, \textit{VisionApp}}, is developed to fuse multimodal inputs derived from vision and radio data, and to perform obstacle-aware gNB mobility control using a Deep Q-Network (DQN). The DQN operates on structured state vectors constructed from fused sensing data and generates discrete gNB mobility actions, which are translated into control messages sent over the E2 interface. Third, a \textbf{3D digital twin environment, \textit{VisionTwin}}, is implemented to emulate realistic mobility conditions, generate synthetic data for training the DQN embedded in \textit{VisionApp}, and enable comprehensive system evaluation.

The rest of this paper is organized as follows. 
Section~\ref{chap:mult} presents the proposed vision-enabled architecture for mobile O-RAN gNBs. 
Section~\ref{chap:xapp} details the design and capabilities of \textit{VisionApp}. 
Section~\ref{chap:cctwin} describes \textit{VisionTwin}, the 3D digital twin environment. 
Section~\ref{chap:eval} outlines the experimental setup and presents the experimental results. 
Section~\ref{chap:discussion} discusses the main advantages and limitations of the system.
Section~\ref{chap:related} presents related work. 
Finally, Section~\ref{chap:conc} concludes the paper and outlines directions for future work.

%% file: 2_VisionRAN.tex
\section{Vision-Enabled O-RAN Architecture} \label{chap:mult}

The state-of-the-art O-RAN architecture provides a standardized E2 interface for near-real-time control of the RAN. This interface is primarily used to exchange network and radio metrics, such as Key Performance Measurements (KPMs) and SNR. While these metrics are essential for monitoring link-level performance, they offer limited understanding of the surrounding physical environment.

We propose two new Service Models (SMs) within \textit{VisionRAN}, a vision-enabled extension of the CONVERGE O-RAN architecture \cite{teixeira2025converge}. In \textit{VisionRAN}, the E2 interface is extended beyond traditional radio and network measurements to support multimodal data exchange. By offloading sensing tasks to edge-deployed E2 agents, the architecture abstracts low-level data acquisition from the network control layer, allowing it to focus on high-level control strategies. xApps operating on the near-RT RIC receive structured information that combines radio measurements, positioning data, and object-level vision features extracted from Computer Vision algorithms. This multimodal integration facilitates perception-driven decision-making.

\subsection{New Vision Service Models for Multimodal Perception}

To enable multimodal perception in \textit{VisionRAN}, we introduce \textbf{two new E2 Service Models (SMs): \texttt{POS} and \texttt{VIS}}. These SMs support the structured delivery of data from E2 Agents to xApps on the near-RT RIC, enabling environment-aware and obstacle-informed mobility control.

The \texttt{POS} SM manages positioning data. It defines indication messages that report to xApps the three-dimensional coordinates and velocities of key network entities, such as the gNB and RGB-D cameras. Additionally, it specifies control messages that allow xApps to manage the position of the gNB. All positional values are expressed within a global coordinate system aligned with the deployment environment.

The \texttt{VIS} SM provides semantic information derived from RGB-D camera inputs. It includes indication messages with object-level detections captured in each camera's field of view, along with position and class labels. This semantic data enhances network situational awareness by allowing xApps to infer LoS conditions, predict obstructions, and adapt control strategies accordingly. The specifications of these SMs are summarized in Tables~\ref{tab:pos1}--\ref{tab:vis2}.

\begin{table}[ht]
\centering
\caption{POS Indication Message Structure}\label{tab:pos1}
\begin{tabular}{l l l}
\toprule
\textbf{Type} & \textbf{Name} & \textbf{Description} \\
\hline
\textit{pos\_stats\_t*} & \textit{pos\_stats} & Pointer to array of \texttt{POS} data entries\\
\textit{uint32\_t} & \textit{len} & Number of entries in the array\\
\textit{int64\_t} & \textit{tstamp} & Timestamp of the report (\textmu s)\\
\bottomrule
\end{tabular}

\vspace{10px}

\caption{POS Data Entry}\label{tab:pos2}
\begin{tabular}{ l l l}
\toprule
\textbf{Type} & \textbf{Name} & \textbf{Description} \\
\hline
\textit{int16\_t} & \textit{id} & Network entity ID \\
\textit{int32\_t} & \textit{x}, \textit{y}, \textit{z} & Cartesian position (cm) \\
\textit{int32\_t} & \textit{vx}, \textit{vy}, \textit{vz} & Velocity vector components (cm/s) \\
\textit{int32\_t} & \textit{theta}, \textit{phi} & Elevation and azimuth angles (rad $\times$ 100) \\
\bottomrule
\end{tabular}

\vspace{10px}

\caption{POS Control Message Structure}\label{tab:pos3}
\begin{tabular}{l l l}
\toprule
\textbf{Type} & \textbf{Name} & \textbf{Description} \\
\hline
\textit{int32\_t} & \textit{x}, \textit{y}, \textit{z} & Target Cartesian coordinates (cm) \\
\textit{int64\_t} & \textit{tstamp} & Timestamp of the control command (\textmu s) \\
\bottomrule
\end{tabular}
\end{table}

\begin{table}[ht]
\centering
\caption{VIS Indication Message Structure} \label{tab:vis1}
\begin{tabular}{l l l}
\toprule
\textbf{Type} & \textbf{Name} & \textbf{Description} \\
\hline
\textit{vis\_stats\_t*} & \textit{vis\_stats} & Pointer to array of \texttt{VIS} data entries \\
\textit{uint32\_t} & \textit{len} & Number of objects detected in the frame \\
\textit{int64\_t} & \textit{tstamp} & Timestamp of the report (\textmu s) \\
\bottomrule
\end{tabular}

\vspace{10px}

\caption{VIS Data Entry (Object Detection)} \label{tab:vis2}
\begin{tabular}{l l l}
\toprule
\textbf{Type} & \textbf{Name} & \textbf{Description} \\
\hline
\textit{int16\_t} & \textit{id} & Unique ID of the tracked object \\
\textit{int32\_t} & \textit{cls} & Class of the object \\
\textit{int32\_t} & \textit{bbx}, \textit{bby} & Bounding box centroid (px) \\
\textit{int32\_t} & \textit{bbw}, \textit{bbh} & Width and height of the bounding box (px) \\
\textit{int32\_t} & \textit{theta}, \textit{phi} & Elevation and azimuth angles (rad $\times$ 100) \\
\textit{int32\_t} & \textit{r} & Distance to object from the camera (cm) \\
\bottomrule
\end{tabular}
\end{table}

Both SMs are designed for seamless integration into existing E2-compliant near-RT RIC systems. One such system is FlexRIC, a lightweight and extensible software development kit (SDK) that enables rapid prototyping of xApps and supports the integration of new E2 SMs within O-RAN infrastructures~\cite{flexric}.

\subsection{E2 Agents for Multimodal Integration}

To facilitate multimodal data transmission and consider external sensing and actuation entities within the O-RAN architecture, this paper employs the concept of the \textbf{\textit{E2 agent}}, as introduced in~\cite{conv_globe}. Each E2 agent adheres to the FlexRIC E2 agent API, ensuring compatibility with the FlexRIC runtime environment, including scheduling, message encoding, and control handling mechanisms. In this architecture, the E2 agent is enhanced to support bidirectional communication with external perception and actuation modules, such as a video camera or a robotic platform transporting the mobile gNB.

Data exchange is performed using lightweight, JSON-formatted messages structured according to the \texttt{POS} and \texttt{VIS} SMs. This enables real-time transmission of indication messages (e.g., positional updates, object detections) and the reception of control commands (e.g., gNB mobility instructions) from xApps. This modular interface supports perception-driven control logic while maintaining full compatibility with existing RIC and gNB implementations. By encapsulating sensing and actuation logic within the E2 agent, the architecture enables scalable RAN expansion without infrastructure changes.

\subsection{\textit{VisionRAN} Architecture}

\begin{figure}[ht]
    \centering
    \includegraphics[width=0.72\linewidth]{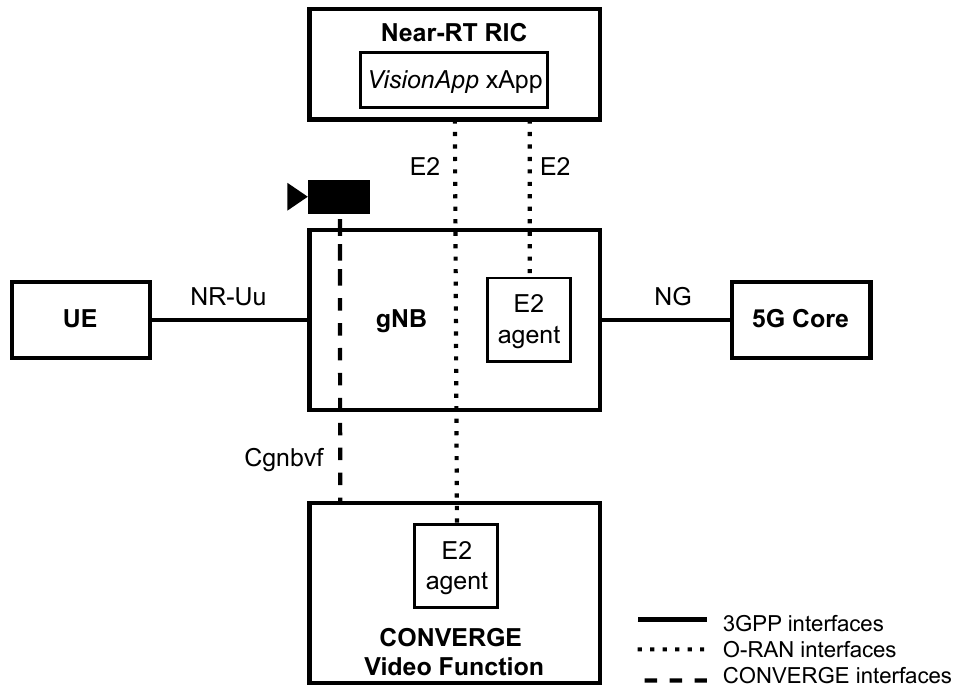}
    \caption{Proposed \textit{VisionRAN} architecture.}
    \label{fig:arch}
\end{figure}

Figure~\ref{fig:arch} illustrates the proposed \textit{VisionRAN} architecture. The CONVERGE Video Function (CVF), developed under the CONVERGE project, is responsible for collecting all the video streams from the cameras. CVF is integrated into the O-RAN control loop via the new E2 agents. CVF applies computer vision algorithms to RGB and depth data streams collected by static and mobile cameras, which are transferred through the Cgnbvf interface. The E2 agents aggregate data, format it into structured \texttt{POS} and \texttt{VIS} indication messages, and transmit them to the near-RT RIC. At the control layer, the \textit{VisionApp} xApp uses this multimodal input to implement perception-aware control strategies, enabling dynamic network behavior that extends beyond traditional radio-based approaches.

This architecture enables the near-RT RIC to consider conventional radio and network metrics, as well as vision-spatial data through the E2 interface, enhancing the detection of UE and obstacle locations for improved gNB positioning control. A detailed system specification is provided in~\cite{converge_spec}. As an extension of the architecture presented in~\cite{conv_globe}, \textit{VisionRAN} proposes two new SMs to support the transmission of positioning and visual information generated by dedicated E2 agents.

%% file: 3_VisionApp.tex
\section{Vision-Aided xApp for gNB Mobility Control} \label{chap:xapp}

The vision-aided xApp, \textit{VisionApp}, implements a \textbf{mobility control policy based on a DQN}, which is a type of deep reinforcement learning algorithm that combines Q-learning with deep neural networks. \textit{VisionApp} aims at repositioning a mobile gNB carried by a robotic platform in order to minimize the duration of non-Line-of-Sight (NLoS) with a UE served by the gNB.

\begin{figure}[ht]
    \centering
    \includegraphics[width=.57\linewidth]{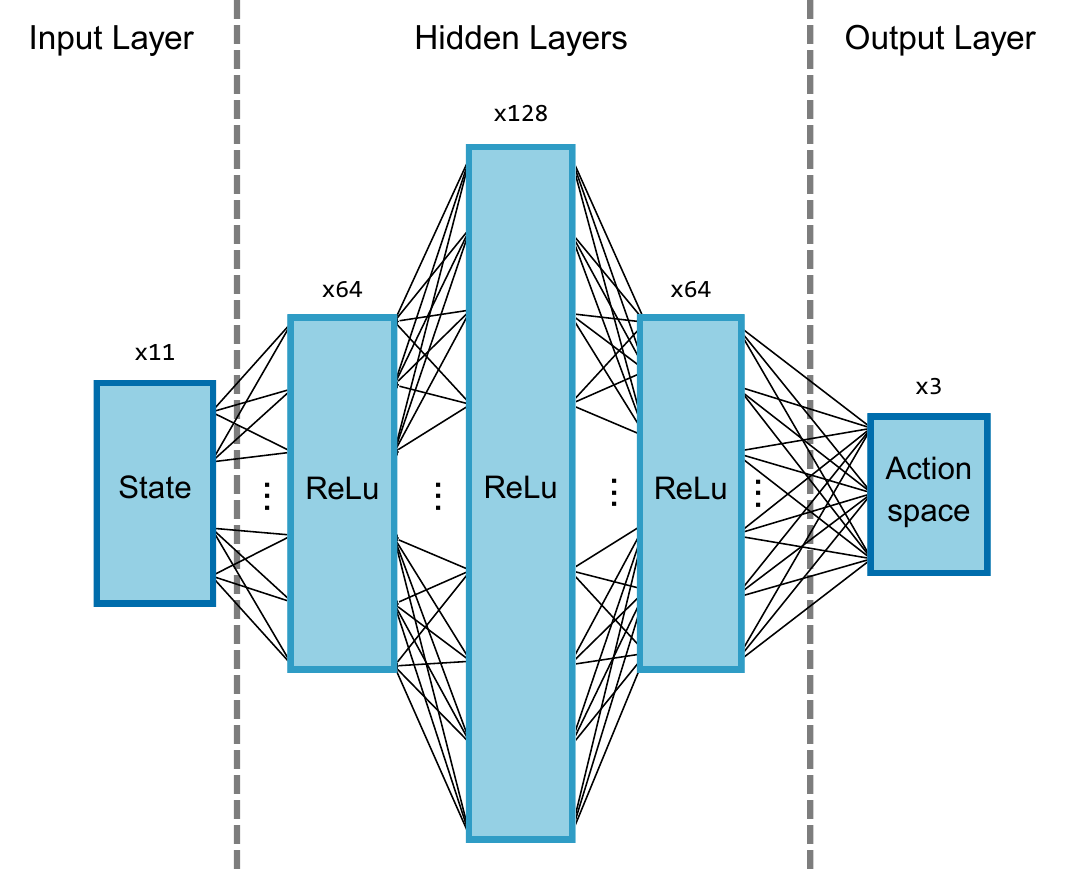}
    \caption{Deep Q-Network architecture.}
    \label{fig:dqn}
\end{figure}

At each decision epoch, the system constructs a structured state vector comprising environmental and mobility-related features. As shown in Table~\ref{tab:inputvec}, the state representation includes the absolute position of the gNB, the relative positions and velocities of the User Equipment (UE) and the obstacle, as well as a binary indicator for the LoS condition. The corresponding action space, summarized in Table~\ref{tab:actionspc}, consists of three discrete actions that govern the velocity of the gNB along the x-axis. The architecture of the trained DQN is illustrated in Fig.~\ref{fig:dqn}.

\begin{table}[ht]
\centering
\caption{State Vector Features} \label{tab:inputvec}
\begin{tabular}{p{2.5cm} p{4.5cm}}
\toprule
\textbf{Feature} & \textbf{Description} \\
\hline
$x_{\text{gnb}}$ & Absolute \(x\)-position of the gNB (m) \\
$x_{\text{gnb-ue}},~y_{\text{gnb-ue}}$ & Relative position: gNB to UE (m) \\
$x_{\text{gnb-obs}},~y_{\text{gnb-obs}}$ & Relative position: gNB to obstacle (m) \\
$v_{x_{\text{gnb}}}$ & gNB velocity along \(x\)-axis (m/s) \\
$v_{x_{\text{ue}}},~v_{y_{\text{ue}}}$ & UE velocity along \(x\), \(y\) (m/s) \\
$v_{x_{\text{obs}}},~v_{y_{\text{obs}}}$ & Obstacle velocity along \(x\), \(y\) (m/s) \\
$L_{\text{status}}$ & LoS status (0 = LoS, 1 = NLoS) \\
\bottomrule
\end{tabular}

\vspace{10px}

\caption{Action Space} \label{tab:actionspc}
\begin{tabular}{p{2.5cm} p{4.5cm}}
\toprule
\textbf{Action} & \textbf{Description} \\
\hline
0 & Maintain current velocity \\
1 & Increase velocity by $\delta$ \\
2 & Decrease velocity by $\delta$ \\
\bottomrule
\end{tabular}
\end{table}

\begin{figure}[ht]
    \centering
    \includegraphics[width=.45\linewidth]{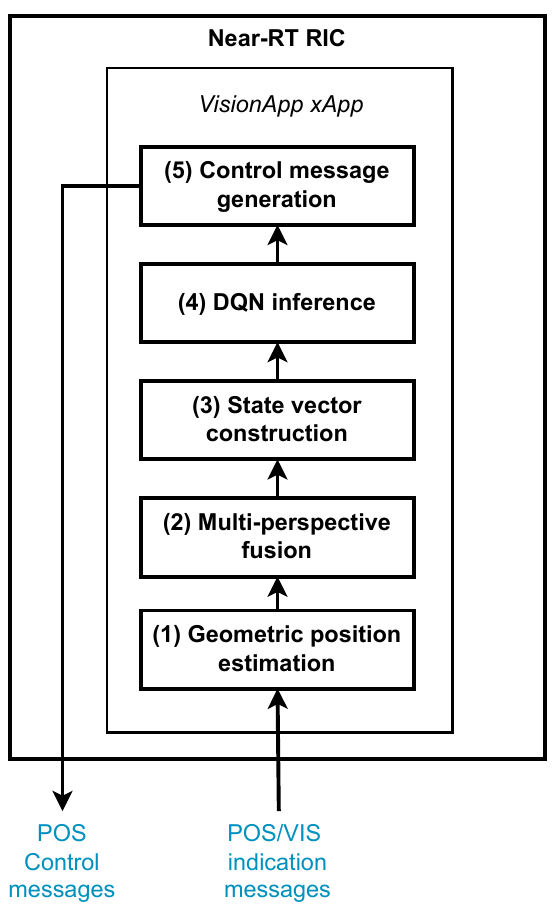}
    \caption{Overview of the algorithmic logic of \textit{VisionApp} within near-RT RIC. \textit{VisionApp} operates with a control interval of 200 ms, which is aligned with the timing constraints of near-RT RIC applications. All processing stages are lightweight and designed to execute within this interval, ensuring practical deployability in real-time O-RAN control loops.}
    \label{fig:xapp}
\end{figure}

During deployment, \textit{VisionApp} operates in real time, receiving multimodal input via E2 interface messages and executing the structured decision loop shown in Fig.~\ref{fig:xapp} every 200\,ms, corresponding to the control interval \(T_{\text{ctrl}}\).

\textit{VisionApp} employs a 5-step process, based on the received \texttt{POS} and \texttt{VIS} messages. The process is described herein and depicted in Fig.~\ref{fig:xapp}. 
In step (1), geometric position estimation converts object detections from polar to Cartesian coordinates, using known reference positions from \texttt{POS} messages. In step (2), multi-perspective fusion combines position estimates from different vision sources, namely the gNB and external cameras within the CONVERGE Video Function to improve UE and obstacle localization by averaging the estimated positions. 
In step (3), state vector construction generates the current state vector based on fused positions and velocities computed via video frame-to-frame differences, as presented in Table~\ref{tab:inputvec}. 
In step (4), DQN inference processes the state vector through the DQN to obtain the optimal control action. 
Finally, in step (5), control message generation computes the new gNB target position $x_{\text{target}}$ as follows: $x_{\text{target}} = x_{\text{gnb}} + v_{\text{new}} \cdot T_{\text{ctrl}}$, where the updated velocity \( v_{\text{new}} \) is derived from the current velocity and the DQN-selected action. The $x_{\text{target}}$ value is embedded into a \texttt{POS} control message and transmitted over the E2 interface to update the gNB's position.

%% file: 4_VisionTwin.tex
\section{Vision-Oriented Digital Twin Environment} \label{chap:cctwin}

To support \textbf{intelligent model training and system emulation}, a vision-oriented digital twin environment, \textit{VisionTwin}, was developed. It provides a controlled, realistic setting for developing and evaluating the \textit{VisionRAN} architecture.

\subsection{Training of Intelligent Models for Vision-Based gNB Control}

In training mode, \textit{VisionTwin} acts as a standalone Reinforcement Learning (RL) platform for optimizing gNB mobility. The simulation engine, \texttt{Environment}, follows the OpenAI Gym API~\cite{Brockman2016}, supporting \texttt{reset()} and \texttt{step()} functions for seamless integration with RL libraries and ensuring reproducibility.

The agent selects actions to adjust the gNB's velocity and position. The environment updates the LoS status and computes path loss to derive a scalar reward. Details on observations and rewards are provided in~\cite{Duarte2025}, which focuses on the training of an RL-based model using synthetic data. It employs an RF model to determine the path loss and considers the same scene layout and control logic as the real network, ensuring direct application in real deployment scenarios.

\subsection{Emulation of gNB Perception and Mobility Control}

\begin{figure}[htbp]
\centering
\includegraphics[width=0.85\linewidth]{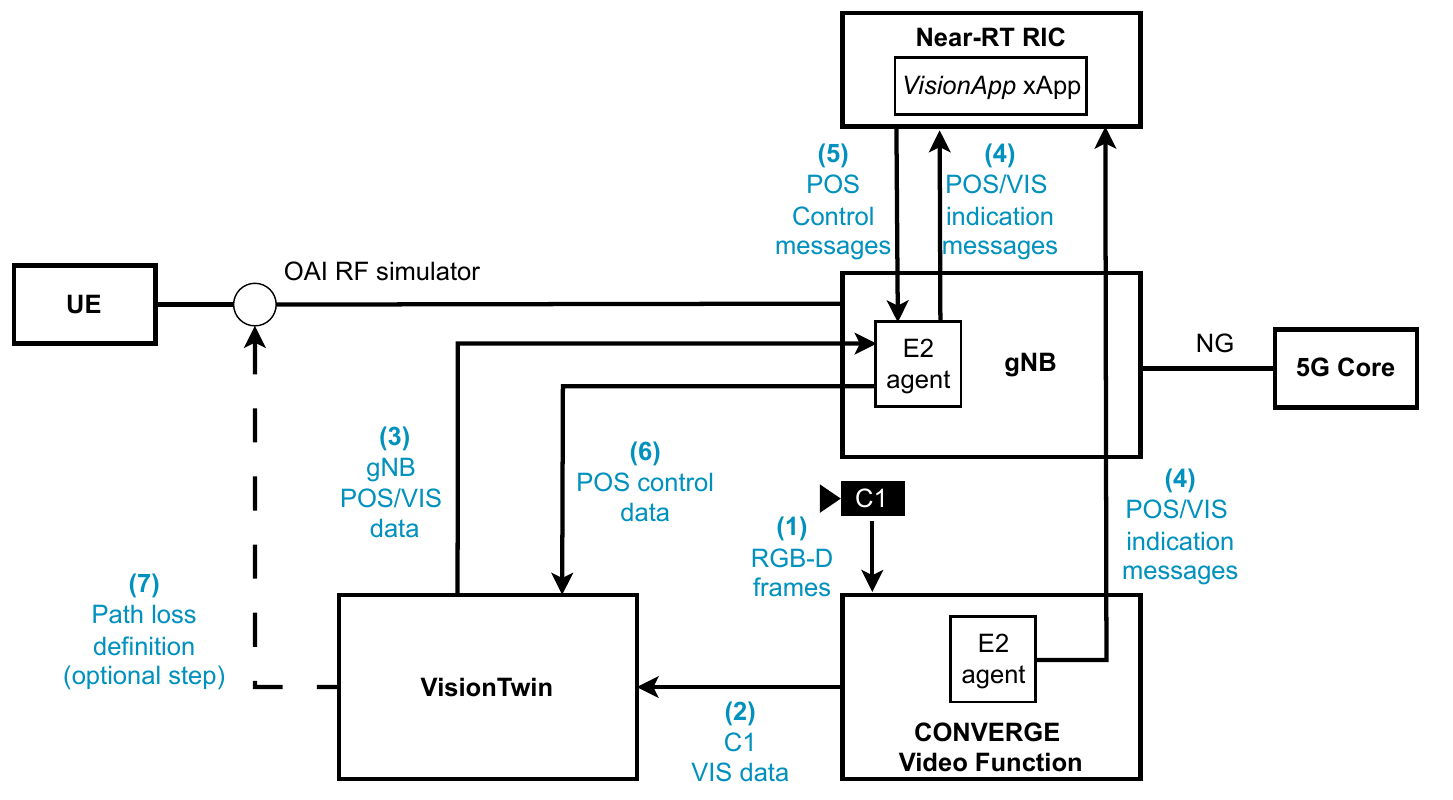}
\caption{Emulation of gNB perception and mobility control using \textit{VisionTwin}.}
\label{fig:val_pipeline}
\end{figure}

The learned policy is validated through closed-loop emulation that replicates real deployment, as depicted in Fig.~\ref{fig:val_pipeline}.

In \textbf{step~(1)}, the CONVERGE reference camera (C1) captures RGB-D video frames. With full visibility, it enables accurate tracking of the UE and obstacles, supporting the perfect-knowledge assumptions used during training. The frames are processed by the CONVERGE Video Function, which extracts object classes, bounding boxes, and polar coordinates, then formats this data into JSON-based \texttt{VIS} messages. 

In \textbf{step~(2)}, vision data from the CONVERGE Video Function is used by \textit{VisionTwin} to reconstruct a dynamic 3D model of the environment, including spatially accurate placements of the UE and obstacle. The current position of the gNB serves to anchor its virtual location within the scene.

In \textbf{step~(3)}, \textit{VisionTwin} simulates the gNB's perception by rendering a visual perspective from the gNB’s viewpoint. It synthesizes this view into \texttt{VIS} and \texttt{POS} data, emulating the output of a real embedded video camera on the gNB during deployment.

In \textbf{step~(4)}, messages from the gNB and CVF E2 agents are sent to the near-RT RIC. The \textit{VisionApp} xApp fuses vision and position data to evaluate LoS and build a semantic state vector.

In \textbf{step~(5)}, this state vector is input into a pretrained Deep Q-Network (DQN), which selects the optimal repositioning action. The resulting command is formatted as a \texttt{POS} control message, which is sent to the gNB's E2 agent.

In \textbf{step~(6)}, the gNB's position is updated in \textit{VisionTwin}, completing the control loop.

Finally, in \textbf{step~(7)}, the path loss of the 5G link between the UE and the gNB is defined by \textit{VisionTwin}. In order to compute the path loss ($P_L$), \textit{VisionTwin} applies:
\begin{equation}
    PL_{\text{dB}} = 20 \log_{10}(d) + A_{\text{obs}} \cdot L_{\text{status}},
\end{equation}
where \(d\) is the distance in meters between the gNB and UE, \(A_{\text{obs}}\) is the obstacle attenuation constant, and \(L_{\text{status}}\) is a binary variable indicating LoS obstruction. The computed path loss is considered by the OpenAirInterface (OAI) RF simulator to model link degradation. Bidirectional traffic is generated using the \texttt{iPerf} tool to evaluate network performance.

%% file: 5_Evaluation.tex
\section{Experimental Evaluation} \label{chap:eval}

To validate the proposed framework and its ability to support intelligent mobility control of a gNB, an experimental setup was developed, which integrates real-world vision sensing data into the \textit{VisionTwin} environment.

\subsection{Physical scenario}

\begin{figure}[ht]
    \centering
    \includegraphics[width=0.7\linewidth]{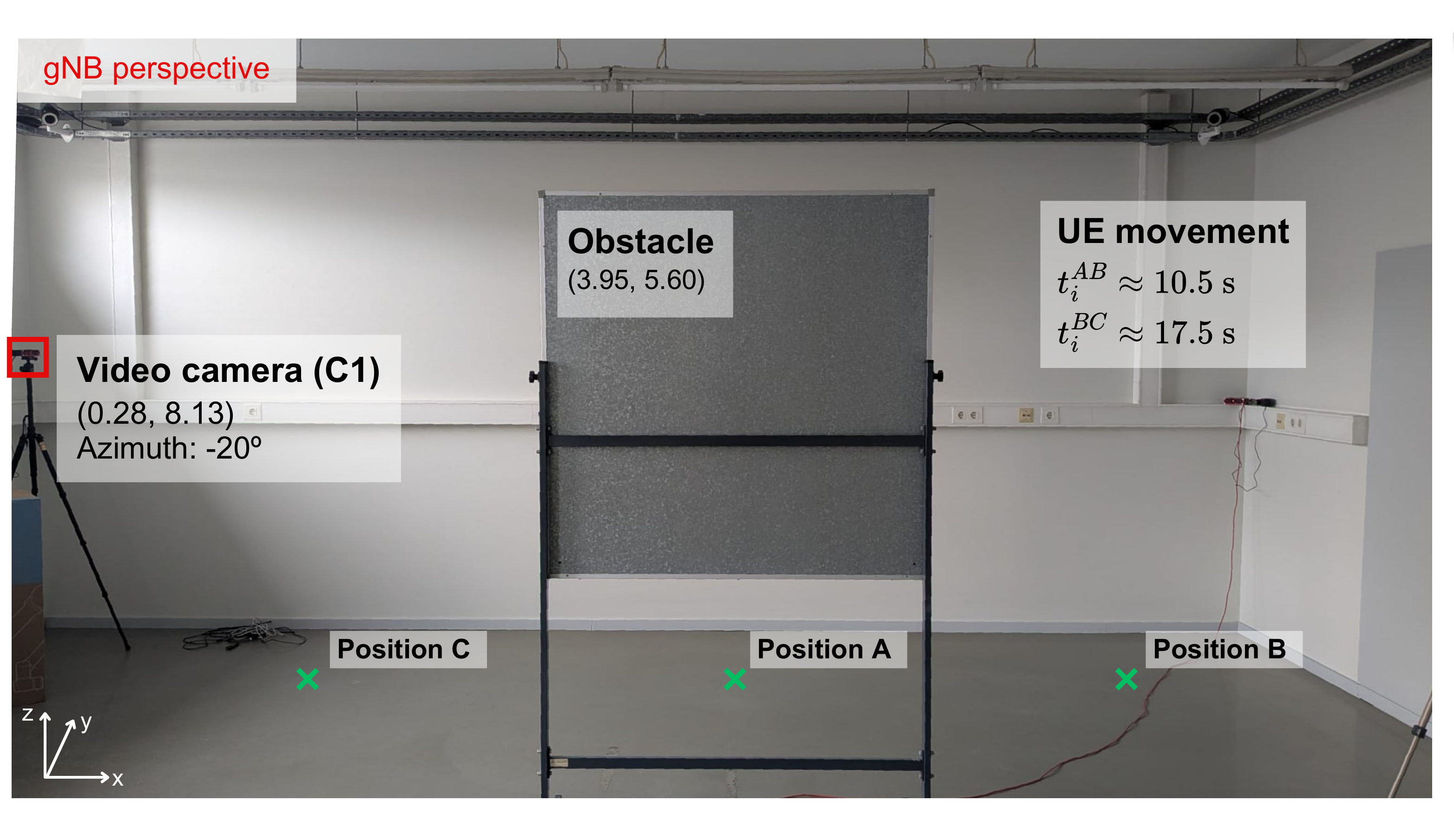}
    \caption{Physical scenario, including an RGB-D video camera and a static obstacle, with annotated reference positions used as ground truth.}
    \vspace{-10px}
    \label{fig:setup}
\end{figure}

The physical scenario consisted of an RGB-D camera (C1), a static obstacle, and a UE. To facilitate validation, the positions and orientations of all entities were manually annotated, as illustrated in Fig.~\ref{fig:setup}.

A 25-second video was recorded at a frame rate of 12\,fps, capturing the relative motion of the UE and obstacle within the video camera’s field of view. The UE, represented by a mobile phone held by a person, initially remained stationary at position A, which was located in a NLoS condition relative to the gNB. At \( t = 10.5\,\text{s} \), the UE was moved to position B, briefly entering a LoS condition. After a short pause, it was relocated to position C at \( t = 17.5\,\text{s} \), where it remained until the end of the sequence. This controlled movement allowed for an evaluation of the system’s responsiveness to dynamically evolving LoS conditions.

\subsection{Positioning accuracy evaluation}

To evaluate obstacle localization accuracy, the estimated obstacle positions \((x_{\text{obs}}^{\text{est}}, y_{\text{obs}}^{\text{est}})\), obtained from \texttt{POS} messages, were compared against manually annotated ground-truth values \((x_{\text{obs}}^{\text{GT}}, y_{\text{obs}}^{\text{GT}})\). The analysis considered several metrics: the \textbf{ground truth}, defined as the manually annotated reference position; the \textbf{mean estimate}, representing the average of all predicted positions for 300 frames captured at 12\,fps over \SI{25}{\second}; the \textbf{maximum and minimum estimates}, indicating the spread of values; and the \textbf{mean deviation from the ground truth}, which is the difference between the ground truth and the mean estimate.

Table~\ref{tab:obstacle-accuracy} summarizes the results for both coordinate axes. The results indicate \textbf{highly accurate} obstacle localization, with mean absolute errors of \SI{8}{\centi\metre} in x and \SI{4}{\centi\metre} in y.

\begin{table}[htbp]
    \centering
    \caption{Accuracy of obstacle position estimation \((x_{\text{obs}}^{\text{est}}, y_{\text{obs}}^{\text{est}})\) compared to ground truth (GT)}
    \label{tab:obstacle-accuracy}
    \begin{tabular}{lcc}
        \toprule
        \textbf{Metric} & \textbf{x (m)} & \textbf{y (m)} \\
        \midrule
        Ground truth \((x_{\text{obs}}^{\text{GT}}, y_{\text{obs}}^{\text{GT}})\)              & 3.95  & 5.60 \\
        Mean estimate \((\bar{x}_{\text{obs}}^{\text{est}}, \bar{y}_{\text{obs}}^{\text{est}})\) & 3.87  & 5.56 \\
        Maximum estimate \((x_{\text{max}}, y_{\text{max}})\)         & 3.99  & 5.64 \\
        Minimum estimate \((x_{\text{min}}, y_{\text{min}})\)         & 3.78  & 5.45 \\
        Mean deviation from GT                              & \textbf{0.08} & \textbf{0.04} \\
        \bottomrule
    \end{tabular}
\end{table}

\begin{figure}[htbp]
    \centering
    \includegraphics[width=0.8\linewidth]{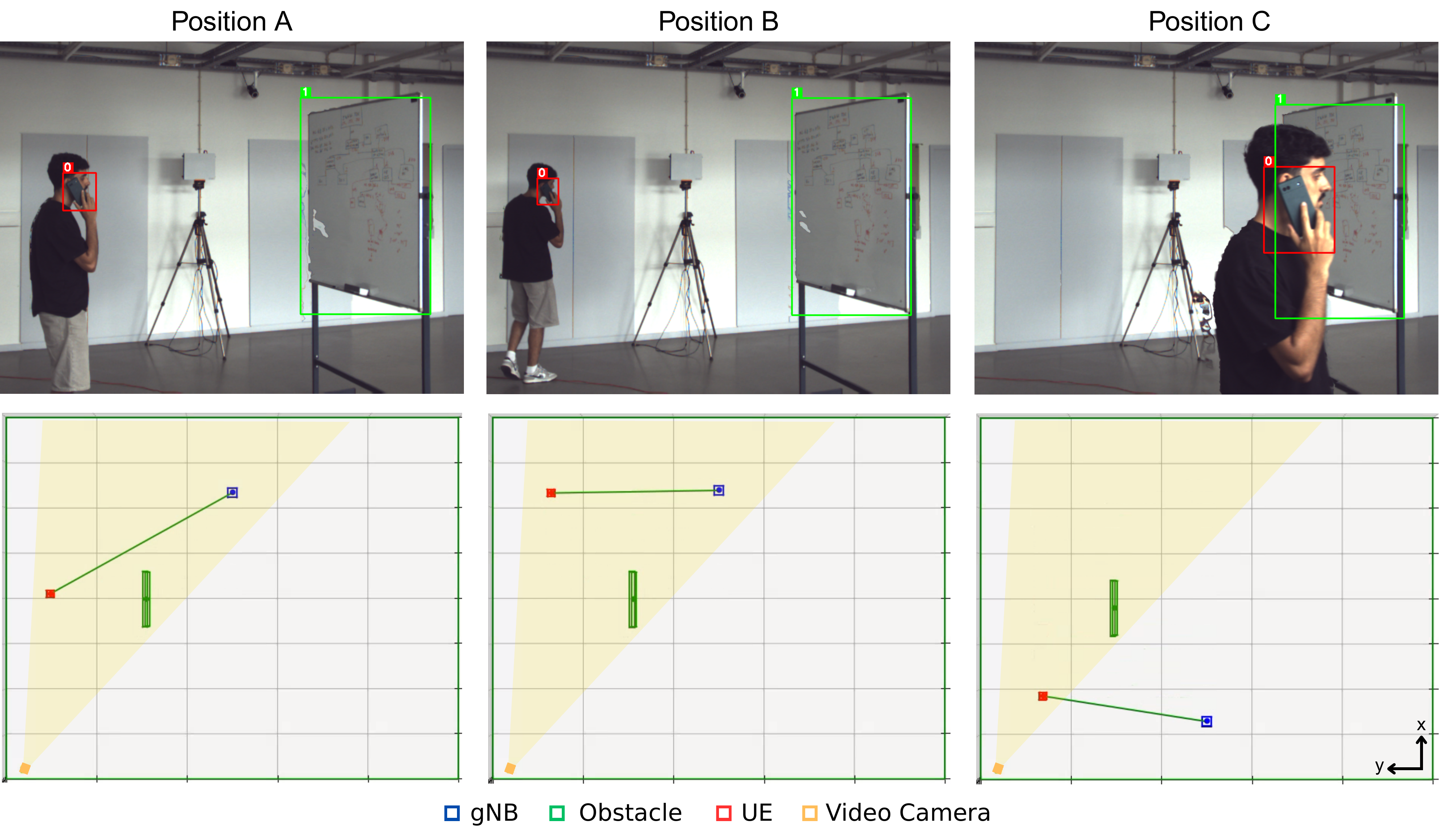}
    \caption{Annotated video frames (top) and corresponding \textit{VisionTwin} emulation states (bottom) for UE positions A, B, and C. While the UE stays in position A, the gNB moves from the center to gain LoS. In position C, the gNB changed its location to maintain LoS with the UE. A video demonstration is available at \cite{video_demo}.}
    \label{fig:frames}
\end{figure}

Using vision-based data, the gNB was dynamically repositioned in response to changing LoS conditions. As illustrated in Fig.~\ref{fig:frames}, the UE was moving across positions A, B, and C, while the gNB adapted its location  to maintain visibility with the UE. These movements were computed by \textit{VisionApp} and executed within \textit{VisionTwin}, enabling direct evaluation of perception-driven mobility control.

\vspace{-10px}
\subsection{Network performance evaluation}

To assess the benefits of gNB mobility control, we compared network performance in two scenarios: one with a static gNB and another with a gNB controlled by \textit{VisionApp}. The obstacle was modeled in the OAI RF simulator for FR1 with an attenuation of 25\,dB, and TCP traffic was generated using \texttt{iPerf}.

\begin{figure}[htbp]
    \centering
    \includegraphics[width=\linewidth]{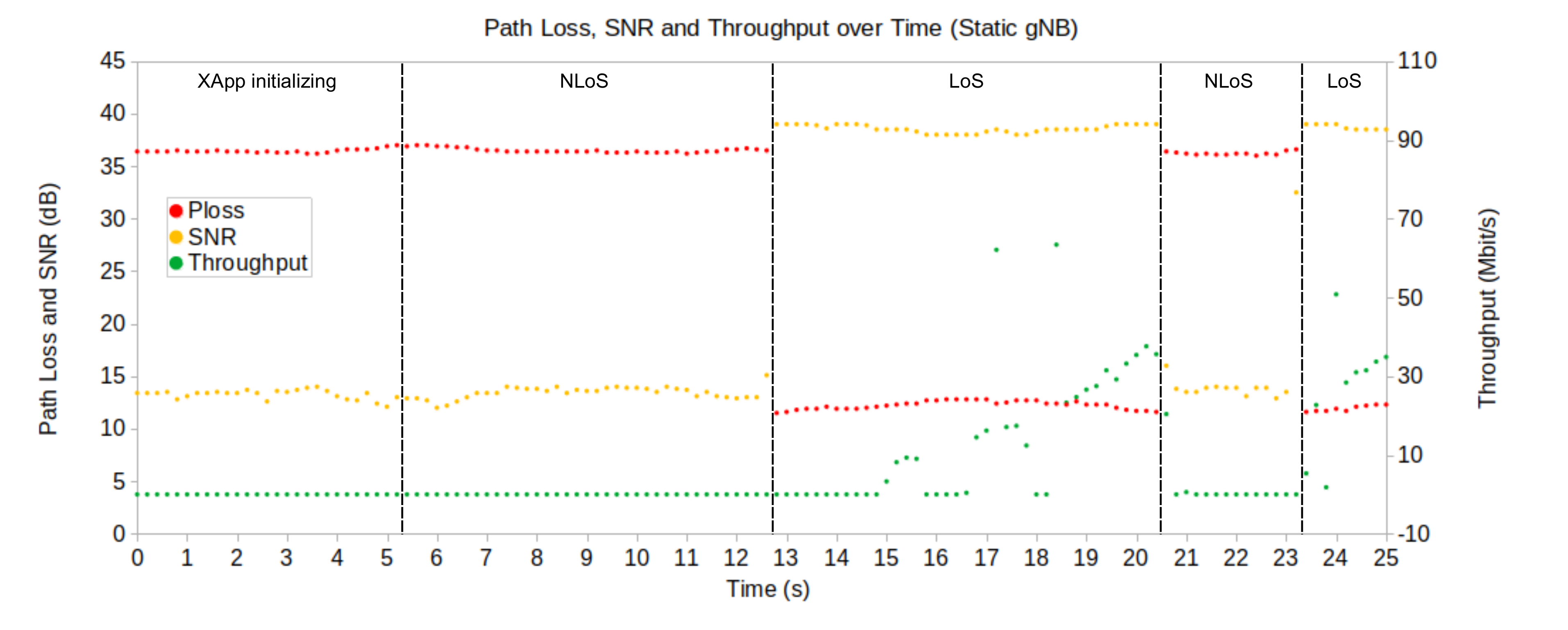}
    \includegraphics[width=\linewidth]{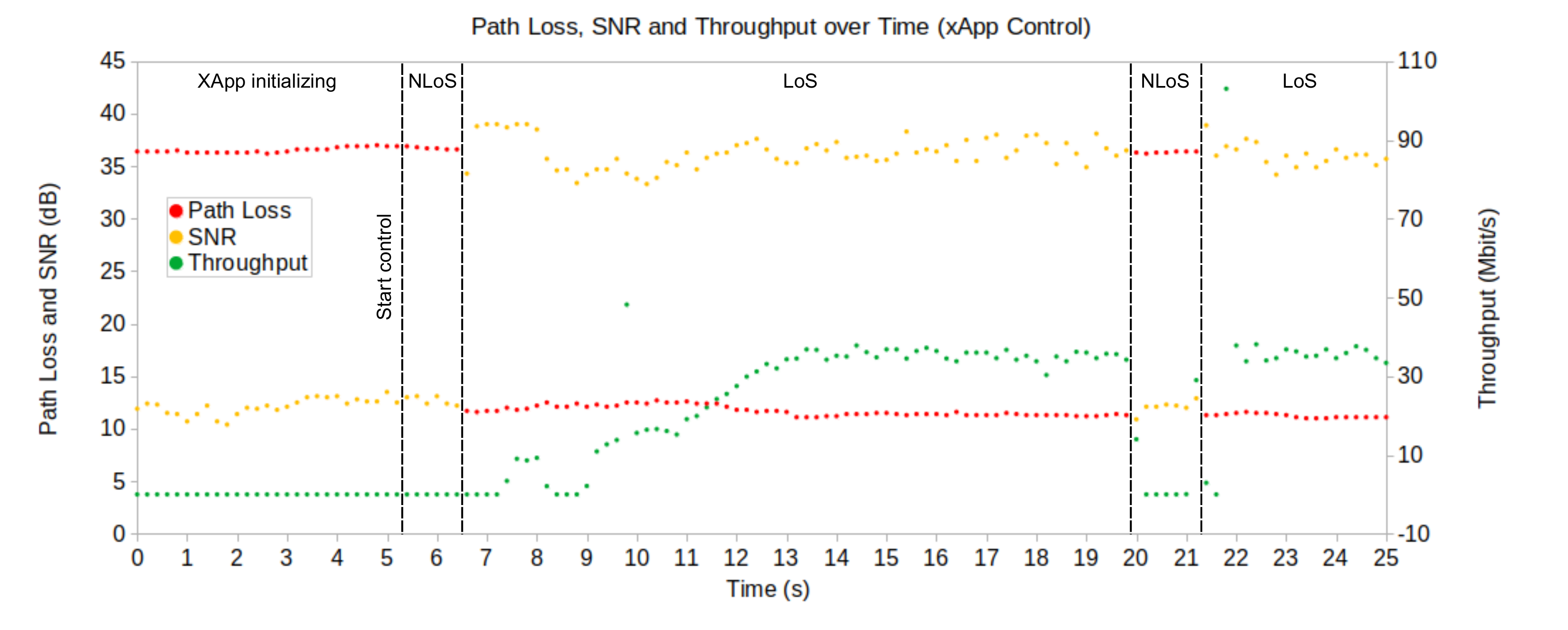}
    \caption{Results showing path loss, SNR, and throughput over time for a static gNB (top figure) and a \textit{VisionApp}-controlled mobile gNB (bottom figure).}
    \vspace{-10px}
    \label{fig:results}
\end{figure}

As shown in Fig.~\ref{fig:results}, a significant reduction in NLoS intervals is observed upon initializing \textit{VisionApp} at \SI{5.4}{\second}. Vertical dashed lines in both subplots indicate the transitions between LoS and NLoS states. In the static gNB scenario (top figure), the gNB remains fixed, resulting in prolonged NLoS intervals. In contrast, when \textit{VisionApp} takes control (bottom figure), the gNB dynamically adjusts its position based on real-time environmental perception. Under \textit{VisionApp} control, NLoS intervals are significantly shorter, leading to a \textbf{75\% reduction in total NLoS duration over the 25-second interval}. This improvement is also reflected in a more stable SNR and consistently higher throughput.


%% file: 6_Discussion.tex
\section{Discussion} \label{chap:discussion}
\vspace{-1px}

The results validate the integration of multimodal perception into an O-RAN-compliant gNB mobility framework. Using real vision data and virtual mobility via \textit{VisionTwin}, the system improved LoS duration and throughput compared to static baselines. A discrete-action DQN enabled real-time, vision-driven control, confirming the feasibility of learning-based gNB mobility within the O-RAN loop.

\textit{VisionTwin} emulated sensing and actuation, interacting with real E2 agents and enabling realistic validation, while adhering to O-RAN standards.
The \textit{VisionApp}, running in FlexRIC, received the vision and position information while controlling the position of the mobile gNB over time.

The current system assumes accurate object detection and depth estimation from the RGB-D camera. Depth errors or misclassifications may degrade environment reconstruction and LoS inference. Future implementations can mitigate this through filtering (e.g., Kalman filters) within the E2 agents before generating POS and VIS messages.

The evaluation focused on a single UE and obstacle to validate the end-to-end framework. While this setup does not assess the generalization of the DQN controller, the architecture and Service Models are scalable to multi-UE and multi-obstacle scenarios, which may require an adapted control module and state representation.

Overall, the proposed framework demonstrates that modular, vision-enhanced control can significantly improve adaptability and performance in 6G networks.

%% file: 7_Related_Work.tex
\vspace{-3px}
\section{Related Work} \label{chap:related}

The evolution of the O-RAN architecture has facilitated near-real-time, learning-based control via standardized interfaces and modular components, such as the near-RT RIC and xApps~\cite{oran, oran-e2}. Open-source near-RT RIC implementations (e.g., FlexRIC~\cite{flexric}) offer extensible support for new SMs, enabling rapid prototyping of RAN control applications.

Most existing network control solutions rely primarily on RF and network-layer metrics, such as KPMs and SNR, which offer limited environmental awareness~\cite{Queiros2024}. To bridge this gap, recent studies have explored the integration of CV and radio sensing for proactive network reconfiguration and semantic environment understanding.
Existing vision-aided wireless networking works can be broadly categorized into:
(i) vision-assisted beam management and blockage prediction without gNB mobility \cite{Alrabeiah2020, Xu2023};
(ii) perception-aware network optimization without O-RAN integration \cite{Kim2024}; and
(iii) O-RAN-based control applications without multimodal perception or mobility support \cite{Queiros2024, Simoes2025, Qazzaz2024, Santos2025}.

Unlike these approaches, the proposed framework uniquely combines standardized O-RAN integration, real vision-based perception, and learning-driven gNB mobility control within a unified architecture.

%% file: 8_Conclusions.tex
\section{Conclusions} \label{chap:conc}

We introduced a vision-based framework for the intelligent control of a mobile gNB operating in dynamic environments. The system integrates three innovative components: \textit{VisionRAN}, which includes two new E2 SMs built for the CONVERGE vision-enabled architecture; \textit{VisionApp}, a DQN-based xApp for multimodal control; 
and \textit{VisionTwin}, a digital twin for training data-based controllers and real-time system emulation.

Experiments with real vision data, as well as emulated gNB and UE, showed up to a 75\% reduction in LoS blockages compared to static gNB deployment, validating the potential of learning-based, perception-aware control in O-RAN systems.

Future work includes extending the framework to scenarios with multiple obstacles and communications nodes, incorporating more sensing modalities, and integrating real gNB mobility platforms that facilitate live RF metric collection within the CONVERGE infrastructure.

%% file: 9_Bibliography.tex
\bibliographystyle{ieeetr}
\bibliography{references}